# Performance of ChatGPT on USMLE: Unlocking the Potential of Large Language Models for AI-Assisted Medical Education

Prabin Sharma[1], Kisan Thapa[1], Prastab Dhakal [2], Mala Deep Upadhaya[3], Dikshya Thapa[1], Santosh Adhikari[4], Salik Ram Khanal[5]

1 University of Massachusetts Boston, USA
2 Texas Tech University, USA
3 Coventry University, Coventry, England
4 MacNeal Hospital, Illinois, USA
5 Center for Precision and Automated Agricultural Systems, Washington State University, Prosser, USA

## Abstract

Artificial intelligence is gaining traction in more ways than ever before. The popularity of language models and AI-based businesses has soared since ChatGPT was made available to the general public via OpenAI. It is becoming increasingly common for people to use ChatGPT both professionally and personally. Considering the widespread use of ChatGPT and the reliance people place on it, this study determined how reliable ChatGPT can be for answering complex medical and clinical questions. Harvard University gross anatomy along with the United States Medical Licensing Examination (USMLE) questionnaire were used to accomplish the objective. The paper evaluated the obtained results using a 2-way ANOVA and posthoc analysis. Both showed systematic covariation between format and prompt. Furthermore, the physician adjudicators independently rated the outcome's accuracy, concordance, and insight. As a result of the analysis, ChatGPT-generated answers were found to be more context-oriented and represented a better model for deductive reasoning than regular Google search results. Furthermore, ChatGPT obtained 58.8% on logical questions and 60% on ethical questions. This means that the ChatGPT is approaching the passing range for logical questions and has crossed the threshold for ethical questions. The paper believes ChatGPT and other language learning models can be invaluable tools for e-learners; however, the study suggests that there is still room to improve their accuracy. In order to improve ChatGPT's performance in the future, further research is needed to better understand how it can answer different types of questions.

## Introduction

The use of Artificial Intelligence (AI) for the human-computer conversation began with the invention of the chatbot. The development of chatbots goes way back in history; with ELIZA being the first chatbot developed by Weizenbaum (Weizenbaum, 1966), successively followed by other noticeable inventions; Artificial Linguistic Internet Computer Entity (ALICE) developed by Wallace (Wallace, 2009), Jabberwacky by Rollo Carpenter (De Angeli et al., 2005), and Mitsuku by Steve Worswick (Abdul-Kader et al., 2015). The AI resides as the backbone of these intelligent agents which can make decisions and responding based on human queries, environment, and experiences which is called model training. The Chatbot is an example of an intelligent agent which uses Natural Language Processing (NLP) to respond like a smart entity when given instruction through text or voice (Khanna et al., 2015). Lexico defines a chatbot as "*A computer program designed to simulate conversation with human users, especially over the Internet*". The NLP uses machine learning algorithms for processing the lexical meaning of words and sentences. These algorithms are mostly based on neural networks and are trained using a big volume and variety of data. The training requires a powerful computing device and takes a very long time to complete. The Chatbots are systems that train on enormous amounts of data for a long time to produce texts, and voice like humans. With the development of powerful deep learning algorithms, the Chatbot jumps to next level with more natural and interactive human-computer conversation.

An American AI research laboratory, OpenAI, released AI-based chatbot called Chat Generative Pre-Trained Transformer (ChatGPT) on November 30, 2022. It is a supervised learning based deep learning model trained using Reinforcement Learning from Human Feedback (Zhang et al., 2023) on fine-tuned GPT-3.5 series which allows asking questions and answers them interactively. The model is trained using billions of texts on azure infrastructure. According to the released documentation of OpenAI (Tom B et al., 2020), the model was trained on almost 570 GB of datasets, which includes books, web pages and other sources (Gratas, 2023). A GPT, Generative Pre-trained Transformer is an autoregressive language model, which uses deep learning transformer models to produce human-like results in text format. ChatGPT uses self-attention mechanisms and a large amount of training data to generate natural language responses to text input in a conversational context.

ChatGPT is one of the largest language models created to date. This model uses a fine-tuned GPT-3.5 model which can perform a variety of tasks, such as question-answering, summarization, and translation. ChatGPT has even been used for generating human-like texts such as stories, poems and even computer code. It has been integrated into various fields like designing, virtual assistants, website chat technology, internet search technology and even messaging apps. It is sometimes criticized to outperform human beings in certain tasks. Currently ChatGPT is made available for developers via API for them to be able to create their own applications, with the help of automation and information generation. ChatGPT has widely impacted companies in technology, education, business services, as well as finance and manufacturing (Zarifhonarvar et al.,2023). As this AI development appears to revolutionize conventional educational procedures, educators' reactions to ChatGPT's extraordinary skills to carry out complex tasks in the field of education have ranged widely (Baidoo-Anu et al., 2023). It is possible to enhance learning and teaching for individuals at all educational levels, including primary, secondary, tertiary, and professional development, by utilizing ChatGPT models. Furthermore, these advanced language models offer a unique opportunity to provide personalized and significant educational experiences because every person has different learning preferences, aptitudes, and needs (Kasneci et al., 2023).

With the wide area of impactful applications, researchers of the similar area show their special interest on ChatGPT related research. Most of the researchers are focused on the evaluation of ChatGPT for answering the questions. Borcji (Borcji et al.) comprehensively describes the ChatGPT failures including reasoning, math, coding, bias, and factual errors while also highlighting the risks, limitations, and societal implications of ChatGPT. They asked ChatGPT questions in several categories and analyzed the generated outputs.

Likewise, Terwiesch (Terwiesch, 2023) experimented ChatGPT using the final exam of Operations Management course, which is one of the core courses of MBA to test the performance of ChatGPT. ChatGPT's performance on an MBA Operations Management final exam revealed both strengths and weaknesses, earning a grade between B and B-. Although it excelled at fundamental operations management and process analysis, it struggled with complex topics and straightforward math. This result highlights how crucial it is to take into account AI's influence on business education, curriculum development, and teaching strategies. One of the noticeable findings of this research is that the response generated by ChatGPT lacks credit and references to the source. The ChatGPT useful case in learning purposes is undeniable.

A study was presented by Wu (Wu, 2023) to perform a comparison of corpus - either human or AI generated - Evaluation and Detection to answer the research question, "How close is ChatGPT to human experts?", by conducting a comprehensive human evaluations and linguistic analysis compared with that of humans.

In the recent years, the students are evaluated online using online exam tools instead of asking for the presidential exam. This trend becomes even more useful during and after Covid-19 pandemic. Most of the recent scientist are focusing on the research about ChatGPT's performance to find out how ChatGPT performs for the university levels examination questions. Similarly, Thurzo et al., (Thurzo et al., 2023) provides brief analysis of use of AI in dentistry, and how it affects the foundations of dental education, including essay, thesis, research paper writing.

One of the most common examinations in USA is United States Medical Licensing Examination (USMLE). A few papers are presented in state of art articles to test the capability of ChatGPT for the USMLE questions. Gilson (Gilson et al., 2022) evaluated the performance of ChatGPT and interpreted the responses on the USMLE Step 1 and Step 2 exams, finding a performance greater than 60% threshold on the step 1 exam making it a viable medical education tool. Medical education tools analysis is an important topic in the education field. So, we dig further in the USMLE exam to evaluate the performance of this Artificial Intelligence language model. Likewise, Kung (Kung et al., 2022) evaluated the ChatGPT performance on the USMLE, where they achieved scores at or near the passing threshold for all three exams without specialized training. Their research suggested ChatGPT potential in medical education and clinical decision-making, though ethical considerations like bias and transparency must be addressed.

Not only medical schools, Choi (Choi, 2023) assessed ChatGPT ability to complete law school exams without human assistance using four real exams from the University of Minnesota Law School. Blind grading revealed that ChatGPT performed at the level of a C+ student, obtaining a passing grade in all courses. There are few concerns researchers raised about the integrity of the exam because of the use of ChatGPT. (Susnjak, 2022) revealed that high-quality text generation with minimal input poses a significant threat to exam integrity, particularly in tertiary education. The researcher also suggested that ChatGPT can generate such answers and suggested some potential countermeasures to the threat of exam integrity. Another concern was raised by (Kasneci et al., 2023), where they discussed the benefits and challenges of using large language models in education, such as creating personalized learning experiences. However issues like potential bias and misuse must be addressed.

Technological advancements in AI and natural language processing have made it possible to develop AI prompt models like ChatGPT that can be used for reasoning on different texts. The Medical Education field is very vast and including books, the internet also serves as a reference for the answers. But can we also depend on Artificial Intelligence for these answers? Can prompt models, especially ChatGPT answer medical questions? If yes, how accurate are the answers? What type of questions does it answer best? Does ChatGPT categorize complexity of questions? Can it be used to train medical personnel as well as the patients? We perform this study to analyze if ChatGPT can complete prevalent tasks related to complex medical and clinical information. In this study, we perform a detailed analysis of ChatGPT on the United States Medical Licensing Examination (USMLE) step one as well as an ethical questionnaire from Harvard University to find the trend in the accuracy and the possibility of using it as an assistant in e-learning.

## MATERIALS AND METHODS
### Data Collection

Based on the objective of the study, the paper incorporated United States Medical Licensing Examination (USMLE) questionnaire (Murphy, 1995). The standard USMLE questions set of steps one consisted of n = 119 total questions and n = 10 short answer questions. To understand the further potential of the ChatGPT, we tested final year questions of Gross Anatomy course of the Harvard University. As the chatbot applications performance depends on how we ask the question, the questions need to be reformatted and screen. In the study, the next step after collecting the questions is to reformat and screen the questions.

ChatGPT only allowed to ask question using text, therefore, the images, charts, and formulae should be removed. The ChatGPT is language model based on the GPT (Generative Pre-trained Transformer) architecture, proving images and concern data is not suitable. 45 among 119 questions consist of images and charts and formulae, therefore, 45 questions are screened. The screened USMLE questions setwas n =74. To analyze the critical reasoning answer of ChatGPT we added 40 questions related to Ethics section of medical exam from Amboss platform (www.amboss.com ).

```
Q. In a sample of 100 individuals, the mean leukocyte count is 7500/mm3, with a standard deviation of 1000/mm3. If the leukocyte
counts in this population follow a normal (gaussian) distribution, approximately 50% of individuals will have which of the
following total leukocyte counts? : (A) 5500-9500/mm3 (B) <6500/mm3 or >8500/mm3 (C) 6500-8500/mm3 (D) <7500/mm3 (E) >9500/mm3
A. C) 6500-8500/mm3
```

*Figure 1 Sample Question of USMLE Exam*

On receiving the questions, we formatted MCQs with no additional prompts or clarifications. Further we grouped questions into subject like Microbiology, System, Pharmacology as per USMLE test coverings.

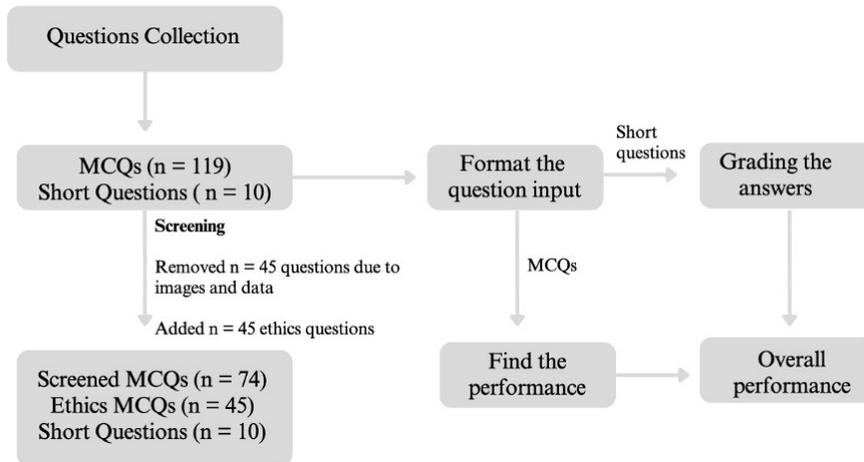

*Figure 2 : Schematic of workflow for sourcing, formatting the questions.*

**Experiments**

ChatGPT responses instantly as it is trained and does not perform online searches unlike other chatbot out there in the internet. Because of this, the ChatGPT is regarded as suitable for handling long-range dependencies and generating coherent and contextually appropriate responses. The ChatGPT provides python API to ask the questions through which the questions can be asked iteratively. For the experiments, a client program is written using ChatGPT API to ask questions listed in a notepad file separating each question by new line and receive the answers from the ChatGPT. And we collected the response of ChatGPT in a separate file. We further proceed to analyze the generated response via ChatGPT.

For the short answer question, as the questions were in pdf format. Firstly, we extracted the questions in the text format similar to MCQ question in the figure 1. The same client program designed for MCQ question was used for the short questions. The response of the ChatGPT was collected in a separate file.

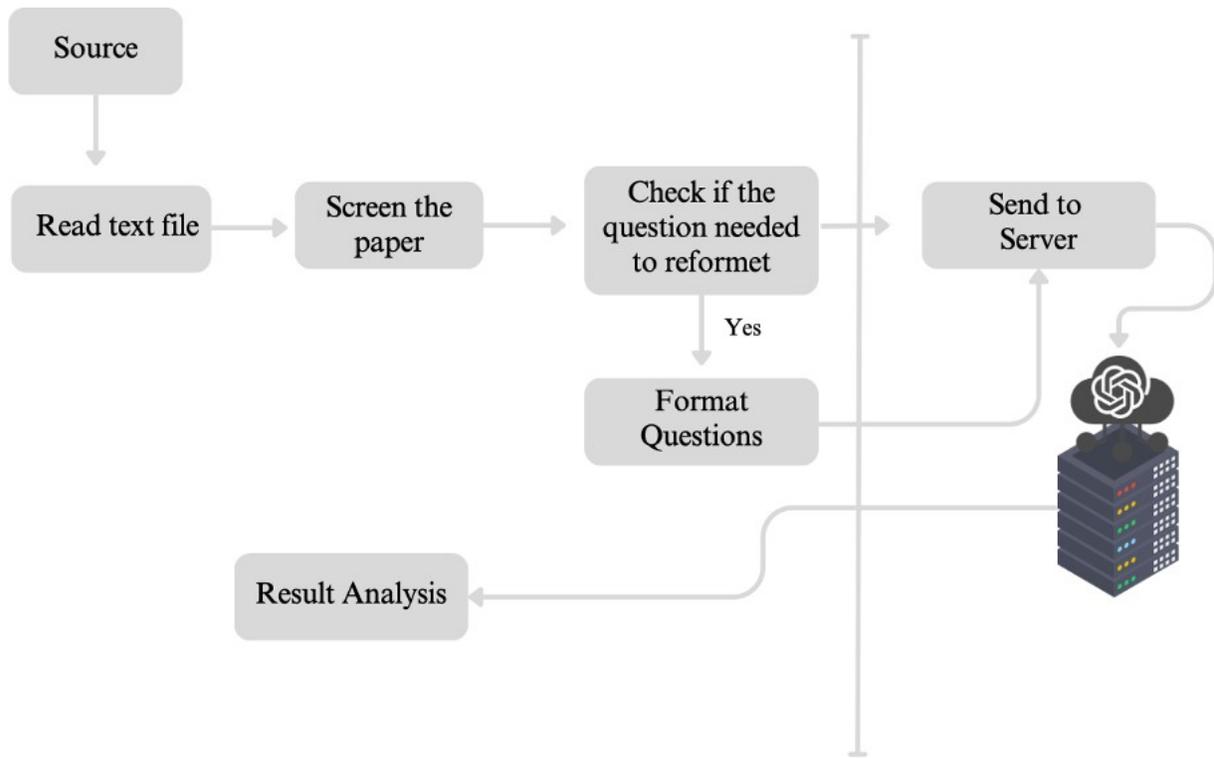

*Figure 3 : Schematic workflow of the source question to ChatGPT API*

**Data Analysis**
A new chat session was created for each entry to handle diversities of the responses and reduce memory retention bias. The results were evaluated using 2-way ANOVA and post *hoc* and obtained the systematic covariation between format and question prompt. On top of this, thus obtained result was independently scored for Accuracy, Concordance, and Insight by physician adjudicators and gave overall performance.

**RESULTS AND DISCUSSION**
It was interesting to see that out of a total of 74 logical questions, 43 of them were correctly answered, 29 of them were answered incorrectly, and 2 of them were not answered at all. In the end, this leads to a rate of accuracy of 58.18% in the overall model. This shows that ChatGPT needs to improve its ability to answer logical questions to be able to serve as a useful analytical tool in the future.

Table 1: Performance test based on the type of question.

| SN | Total Questions | Not Answer | Correct Answers | Wrong Answer | Accuracy (%) |
|---|---|---|---|---|---|
| **Logical Questions** | 74 | 2 | 43 | 29 | 58.18 |
| **Critical reasoning (Ethics part)** | 45 | 0 | 25 | 20 | 60 |

It is worth noting that some questions were left unanswered by ChatGPT. Specifically, there were four questions that were not answered in the Logical Questions section, and one question each in the Biostatistics and Immunology sections. It may also be worth investigating why ChatGPT was not able to answer these questions and whether there are any patterns or commonalities among them.

ChatGPT gave wrong answers for a total of 38 questions. It may be worth investigating whether there are any patterns or commonalities among these questions, such as certain types of questions or subjects where ChatGPT struggles to provide accurate answers.

As far as the critical reasoning questions related to ethics are concerned, among 45 questions, 25 were answered correctly, 20 were answered incorrectly, and no question was not answered at all. It appeared that ChatGPT has improved its ability to handle critical reasoning questions compared to itself answering logical questions. However, there is still room for improvement in the ability to handle critical reasoning questions with ChatGPT. In course wise evaluation, the ChatGPT provided high accuracy on Microbiology and immunology with 80% and 75% respectively and the minimum accuracy rate of 20% was obtained for Anatomy. The figure 3 shows the bar diagram of the accuracy and the performance on each course.

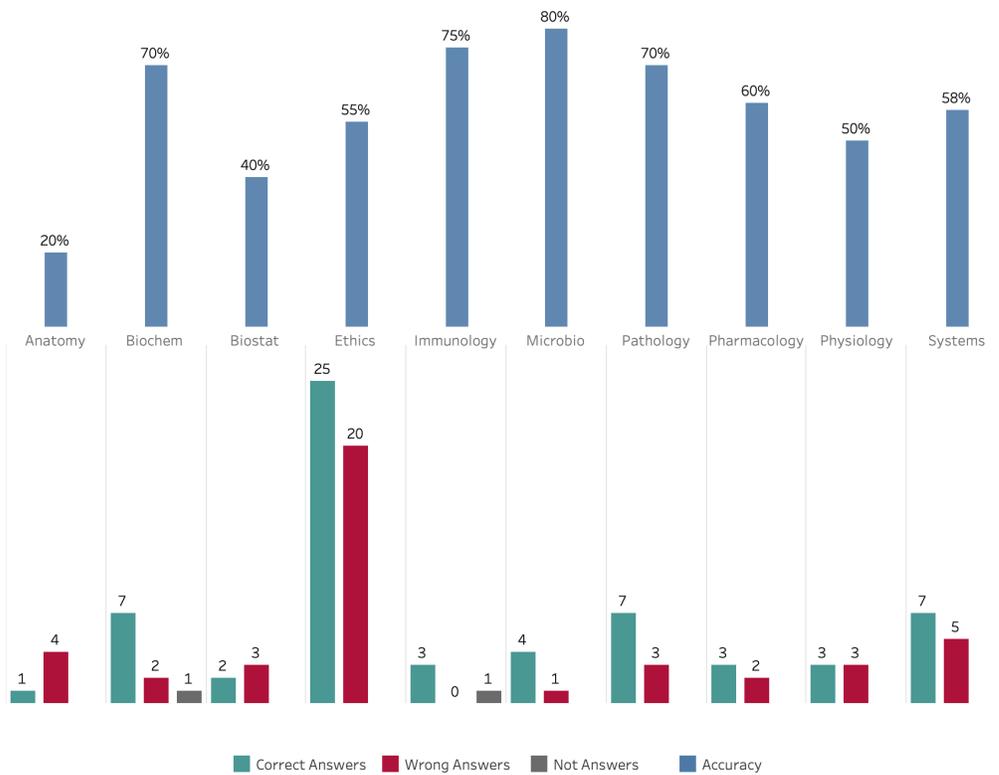

*Figure 4 : Performance test based on the type of subject.*

To identify any patterns or trends in the performance of the Chat GPT across different course that could be derived from the data presented in the table, we conducted a statistical analysis of the data presented in the table. Firstly, the mean accuracy rate across all subjects was calculated to be 59.5%, indicating that on average ChatGPT accurately answered slightly more than half of the questions posed.

Post-hoc tests were performed after an initial statistical analysis, such as ANOVA, to determine which pairs of groups differ significantly from each other. In this case, the accuracy rates across subjects were analyzed using a one-way ANOVA. This analysis indicates a significant difference in accuracy rates across subjects ($F(9, 1) = 9.10$, $p < 0.05$). The diagonal entries are all equal to "-" because they represent the comparison of a subject with itself which is not meaningful.

To evaluate the course wise association, a post-hoc pairwise comparisons using a Tukey HSD test was used. The results of these tests are presented in Table 2.

Table 2: Post-hoc pairwise comparisons for accuracy rates of nine courses.

|  | Biostat | Pathology | Microbio | Systems | Biochem | Ethics | Anatomy | Immunology | Pharmacology | Physiology |
|---|---|---|---|---|---|---|---|---|---|---|
| Biostat | - | 0.119 | 0.001 | 0.660 | 0.119 | 0.847 | 0.001 | 0.311 | 0.716 | 0.973 |
| Pathology | 0.119 | - | 0.003 | 0.052 | 0.735 | 0.233 | 0.001 | 0.039 | 0.512 | 0.254 |
| Microbio | 0.001 | 0.003 | - | 0.019 | 0.003 | 0.843 | 0.001 | 0.001 | 0.141 | 0.035 |
| Systems | 0.660 | 0.052 | 0.019 | - | 0.052 | 0.827 | 0.019 | 0.535 | 0.981 | 0.525 |
| Biochem | 0.119 | 0.735 | 0.003 | 0.052 | - | 0.735 | 0.001 | 0.039 | 0.512 | 0.254 |
| Ethics | 0.847 | 0.233 | 0.843 | 0.827 | 0.735 | - | 0.843 | 0.843 | 0.993 | 0.997 |
| Anatomy | 0.001 | 0.001 | 0.001 | 0.019 | 0.001 | 0.843 | - | 0.001 | 0.141 | 0.035 |
| Immunology | 0.311 | 0.039 | 0.001 | 0.535 | 0.039 | 0.843 | 0.001 | - | 0.648 | 0.175 |
| Pharmacology | 0.716 | 0.512 | 0.141 | 0.981 | 0.512 | 0.993 | 0.141 | 0.648 | - | 0.827 |
| Physiology | 0.973 | 0.254 | 0.035 | 0.525 | 0.254 | 0.997 | 0.035 | 0.175 | 0.827 | - |

This means that the difference between the mean accuracy rate of "Biostat" and "Pathology" is not statistically significant at the 0.05 level since the p-value (0.119) is greater than 0.05. Similarly, the post hoc accuracy between Microbiology and Pharmacology is 0.141 demonstrating it's not statistically significant. Which inferred that the difference between the mean accuracy rate of "Microbio" and "Pharmacology" is not statistically significant at the 0.05 level, since the p-value (0.141) is greater than 0.05. On the other hand, the entry in the row labeled "Microbio" and the column labeled "Physiology" is "0.035", which means that the difference between the mean accuracy rate of "Microbio" and "Physiology" is statistically significant at the 0.05 level, since the p-value (0.035) is less than 0.05.

Table 3: Performance for the short questions.

| SN | Physician I | Physician II | Physician III |
|---|---|---|---|
| Not Answers | | | |
| Grade | 100% | 100% | 100% |
| CGPA | A | A | A |

We wanted to check if ChatGPT works well on short answer questions. We asked ChatGPT ten short answer questions which was asked in Harvard Medical exam of Gross Anatomy of First Year Final exam. We firstly asked ChatGPT those questions and extracted those answers in a word file. Secondly, we asked three physicians to evaluate those answers. Based on their evaluation we found out the ChatGPT scored an A from the evaluation of all three physicians.

Table 4: Google VS ChatGPT

| SN | Google | ChatGPT |
|---|---|---|
| Correct answer | 12 | 15 |
| Wrong answer | 5 | 4 |
| Not Answered | 3 | 1 |

For Google, out of 20 total questions (12 correct, 5 wrong, 3 unanswered), the overall accuracy rate would be 60%. For ChatGPT, out of 20 total questions (15 answered, 4 correct, 1 wrong), the overall accuracy rate would be 80%. Some bias can occur which can impact the performance of both systems. Sample bias: The questions asked may not be representative of the full range of topics and difficulty levels that are covered by the subject matter. For example, if the questions were all on a particular subtopic within a larger subject area, then the performance of Google and ChatGPT on those questions may not be indicative of their overall performance on the course.

**DISCUSSION**

The purpose of this study was to test the performance of ChatGPT or answering capability for the complex medical and clinical related questions. To assess ChatGPT answering capabilities for biomedical and clinical questions of standardized complexity and difficulty, a number of tests have been conducted to determine its performance characteristics, including the United States Medical Licensing Examination (USMLE) step one as well as an ethical questionnaire from Harvard University. We found two major themes emerging from the analysis: (1) the increasing accuracy of ChatGPT; and (2) the possibility of using this AI in a way to assist e-learning. USMLE pass threshold is not constant, it changes every year and on general level it should be 60%. On our study we see ChatGPT obtained 58.8% on logical questions and 60% on ethical one. With this we see the ChatGPT is approaching the passing range for logical question and has touched the threshold for ethical one. The result from this study obtained accuracy quite high compared to GPT LLM (Liévin et al., 2023) which has achieved 46% accuracy with zero prompting and extensive prompt tuning got to 50%. Also, when comparing ChatGPT for MCQs based exam it achieved 53.8% which is 5% below to the accuracy the study achieved. The possibility of using this AI in a way to assist e-learning. Considering the results of the study, it was found that the AI-generated answers were more context-oriented and a better role model for a deductive reasoning process compared to the Google search results. In approximately 90% of the outcomes generated by ChatGPT, there was at least one meaningful insight that was present in at least one of the responses, suggesting that ChatGPT could be a useful platform for enhancing the delivery of e-learning. In a study conducted by (Choi et al., 2023), ChatGPT was found to achieve an average grade of a C+ in all four courses, achieving a low but passing grade for each. It's clear from the above that using ChatGPT the student should be able to get a passing score in the exam.

Anatomy was the subject where our accuracy was a minimum of 20% and Biostatistics was a minimum of 40%, the rest of the subjects were above 50%. We can also conclude that with ChatGPT, passing marks are possible. After all, every student's first goal should be to pass the examination to graduate.

Overall, we found the results of these studies suggest that despite the ability of ChatGPT to answer both logical and critical reasoning questions related to ethics, there still appears to be room for improvement in terms of its accuracy as well. There is a pressing need for further research to better understand how ChatGPT can be used to answer different types of questions and to identify strategies that can improve its performance in the future.

**Conclusion**

In conclusion, this study aimed to assess the reliability of ChatGPT in answering complex medical and clinical questions. The results indicate that ChatGPT shows potential as a valuable tool for e-learners, but there is still room for improvement in terms of its accuracy. The analysis revealed that ChatGPT-generated answers were more context-oriented and demonstrated better deductive reasoning abilities compared to regular Google search results. However, the study found that ChatGPT needs to enhance its performance in answering logical questions to become a useful analytical tool. The study also highlighted the importance of further research to explore how ChatGPT can effectively answer different question types and to identify strategies for improving its performance. Overall, ChatGPT's performance in this study suggests its potential to assist in e-learning, but ongoing advancements are needed to optimize its accuracy and broaden its applications in the field of medicine and clinical practice.